\documentclass[prb,aps]{revtex4-2}
\usepackage[utf8]{inputenc}
\usepackage{amsmath, amssymb, graphicx, appendix, comment}
\usepackage{url, booktabs, bm, stackrel, amssymb}
\usepackage[colorlinks=true, citecolor=blue, urlcolor=blue]{hyperref}
\setcitestyle{super,open={},close={}}

\begin{document}

\title{Emergence of Unpinned Dirac Points in Defected Photonic Crystals}

\author{Chang-Hwan Yi}
\affiliation{Center for Theoretical Physics of Complex Systems, Institute for Basic Science, Daejeon 34126, Republic of Korea}

\author{Kyoung-Min Kim}
\email{kyoungmin.kim@apctp.org}
\affiliation{Asia Pacific Center for Theoretical Physics, Pohang, Gyeongbuk, 37673, Republic of Korea}
\affiliation{Department of Physics, Pohang University of Science and Technology, Pohang, Gyeongbuk 37673, Korea}
\affiliation{Center for Theoretical Physics of Complex Systems, Institute for Basic Science, Daejeon 34126, Republic of Korea}

\date{\today}

\begin{abstract} 
Unpinned Dirac points (DPs) are nodal point degeneracies that occur at generic momentum points lacking high symmetry, often exhibiting characteristics typically forbidden by symmetry. While this phenomenon has been observed in solid-state materials, the formation of unpinned DPs in photonic crystals has not been extensively studied. This study presents a novel approach to achieving unpinned DPs in two-dimensional ``defected photonic crystals"—photonic superlattices created by modifying the size or refractive index of dielectric disks at specific sites. The defected photonic crystal hosts multiple unpinned DPs through accidental band crossings induced by rotational symmetry breaking across the nodal line degeneracies prevalent in the superlattice. Furthermore, the number of unpinned DPs, along with their positions and anisotropic dispersions, can be manipulated by adjusting the size or refractive index of the dielectric disks. This study pioneers the exploration of defected photonic crystals, proposing them as effective means of achieving unpinned DPs and facilitating the investigation of novel relativistic photonic wave phenomena with significant flexibility and control.
\end{abstract}

\maketitle


\section{Introduction}

The search and design of new photonic crystal (PhC) structures exhibiting nodal point degeneracies have become a major research focus in photonics \cite{PhysRevLett.100.013904, PhysRevA.78.033834, PhysRevA.75.063813, PhysRevLett.100.113903, Wang2009, DIEM20102990, Huang2011, doi:10.1073/pnas.1207335109, PhysRevB.86.035141, Khanikaev2013, Rechtsman2013, Lu2013, Li:13, Lu2014, https://doi.org/10.1002/lpor.201300186, doi:10.1126/science.aaa9273, Bravo-Abad_2015, He2015, PhysRevA.93.061801, Chen2016, Khanikaev2017, Noh2017, Slobozhanyuk2017, Yang:17, Wang2017, PhysRevLett.119.213901, PhysRevA.96.013813, PhysRevLett.119.113901, PhysRevB.96.075438, doi:10.1126/science.aaq1221, https://doi.org/10.1002/lpor.201700271, PhysRevLett.121.024301, PhysRevLett.121.263901, Wang2019, Liu2020, Li_2020, PhysRevA.101.033822, Dong2021, Navarro-Barón2021, Chen:21, Chu:21, https://doi.org/10.1002/lpor.202100452, Wang:22, PhysRevB.106.235303, doi:10.1126/sciadv.abq4243, Pan2023, Nakatsugawa:24, FENG2024114719}. These nodal points, known as Dirac points (DPs) and Weyl points (WPs), are characterized by conical wavevector-frequency dispersions, which mimic the relativistic energy dispersion of Weyl fermions. This analogy enables the exploration of various relativistic quantum phenomena governed by the Dirac equation, such as Zitterbewegung \cite{PhysRevLett.100.113903} and Klein tunneling \cite{PhysRevA.96.013813, Nakatsugawa:24}, alongside novel optical phenomena, including pseudodiffusive transport \cite{PhysRevA.75.063813, DIEM20102990}, zero-refractive-index metamaterials \cite{Huang2011, Dong2021}, synthetic magnetic fields for photons \cite{Rechtsman2013}, and anomalous refraction \cite{Wang2017, Wang:22}. Additionally, the non-trivial band topology associated with these nodal points, which can be achieved through time-reversal or inversion symmetry breaking, leads to symmetry-protected localized modes and unidirectional waveguides that have numerous photonic applications \cite{Khanikaev2013, Lu2014, Khanikaev2017}, including backscattering-immune transport \cite{Wang2009}. These notable benefits have led to intensified efforts in searching for novel PhC structures over the past decade, culminating in the realization of DPs and WPs across diverse experimental settings \cite{Wang2009, Lu2013, Rechtsman2013, Khanikaev2013, Lu2014, Noh2017, Khanikaev2017, Liu2020, Slobozhanyuk2017, doi:10.1126/sciadv.abq4243, https://doi.org/10.1002/lpor.201700271, PhysRevLett.121.024301, doi:10.1126/science.aaa9273, Chen2016, doi:10.1126/science.aaq1221, https://doi.org/10.1002/lpor.201700271, https://doi.org/10.1002/lpor.202100452, Pan2023}.

In most two-dimensional (2D) periodic systems, point group symmetry constrains nodal degeneracies at high-symmetry points in reciprocal space; for instance, two DPs are pinned at the corners of the hexagonal Brillouin zone due to the C\textsubscript{6v} symmetry in a honeycomb lattice \cite{PhysRevLett.100.013904, PhysRevA.78.033834, PhysRevA.75.063813, PhysRevLett.100.113903, DIEM20102990, doi:10.1073/pnas.1207335109, PhysRevB.86.035141, Li:13, https://doi.org/10.1002/lpor.201300186, PhysRevA.96.013813, PhysRevLett.121.263901, Navarro-Barón2021}. However, such pinned nodal points are not the only possibility, as accidental band crossings can lead to nodal points at generic momentum locations \cite{doi:10.1142/S2010324716400038}. In solid-state systems, this potential has been explored in a puckered honeycomb structure for the first time \cite{Lu2016}, and relevant unpinned WPs have recently been experimentally realized in related materials \cite{Lu2022}. Additionally, some optical metasurfaces and non-regular arrays of dielectric disks exhibit signatures of unpinned nodal points \cite{He2015, PhysRevLett.119.113901, PhysRevLett.121.024301, PhysRevB.96.075438, Li_2020, Chu:21}. However, achieving unpinned nodal points in PhCs has not been extensively addressed and remains an open challenge. Notably, unpinned nodal points often exhibit anisotropic or nonlinear dispersions \cite{Wu:14}, as well as mobility in reciprocal space in response to external stimuli \cite{Lu2016, Lu2022, doi:10.1142/S2010324716400038}, which are typically forbidden by symmetry for pinned nodal points. Consequently, achieving unpinned nodal points is crucial for exploring the relativistic quantum phenomena and photonic applications that require flexibility and control over these features.

In this study, we investigate a novel 2D PhC structure, termed a ``defected PhC," where lattice periodicity and point group symmetry are disrupted by modifying the size or refractive index of dielectric disks at specific sites that serve as ``defects" [Fig.~\ref{fig:dirac_point}(a)]. While similar superlattice structures have been previously explored in 2D solid-state materials, such as graphene porous nanomeshes \cite{Bai2010, C4NR04584J}, our study represents the first exploration of this concept in PhCs. A photonic superlattice devoid of defects exhibits nodal line degeneracies across high-symmetry lines. However, these degeneracies can be lifted by defects, resulting in the emergence of multiple unpinned DPs [Fig.~\ref{fig:dirac_point}(b)]. The formation of unpinned DPs is attributed to accidental band crossings induced by the breaking of rotational symmetry, while their abundance is associated with the numerous nodal line degeneracies present in the superlattice. Additionally, we demonstrate that the number of unpinned DPs, as well as their positions and anisotropic dispersions, can be controlled by adjusting the size or refractive index of the dielectric disks. Our findings suggest that breaking symmetry through defects, in conjunction with prevalent nodal line degeneracies, is an effective strategy for realizing unpinned DPs in 2D PhCs, thereby facilitating the exploration of novel relativistic photonic wave phenomena with significant flexibility and control.


\section{Result}

\subsection{Emergence of unpinned Dirac points}

\begin{figure*}[t!]
    \centering
    \includegraphics[width=\textwidth]{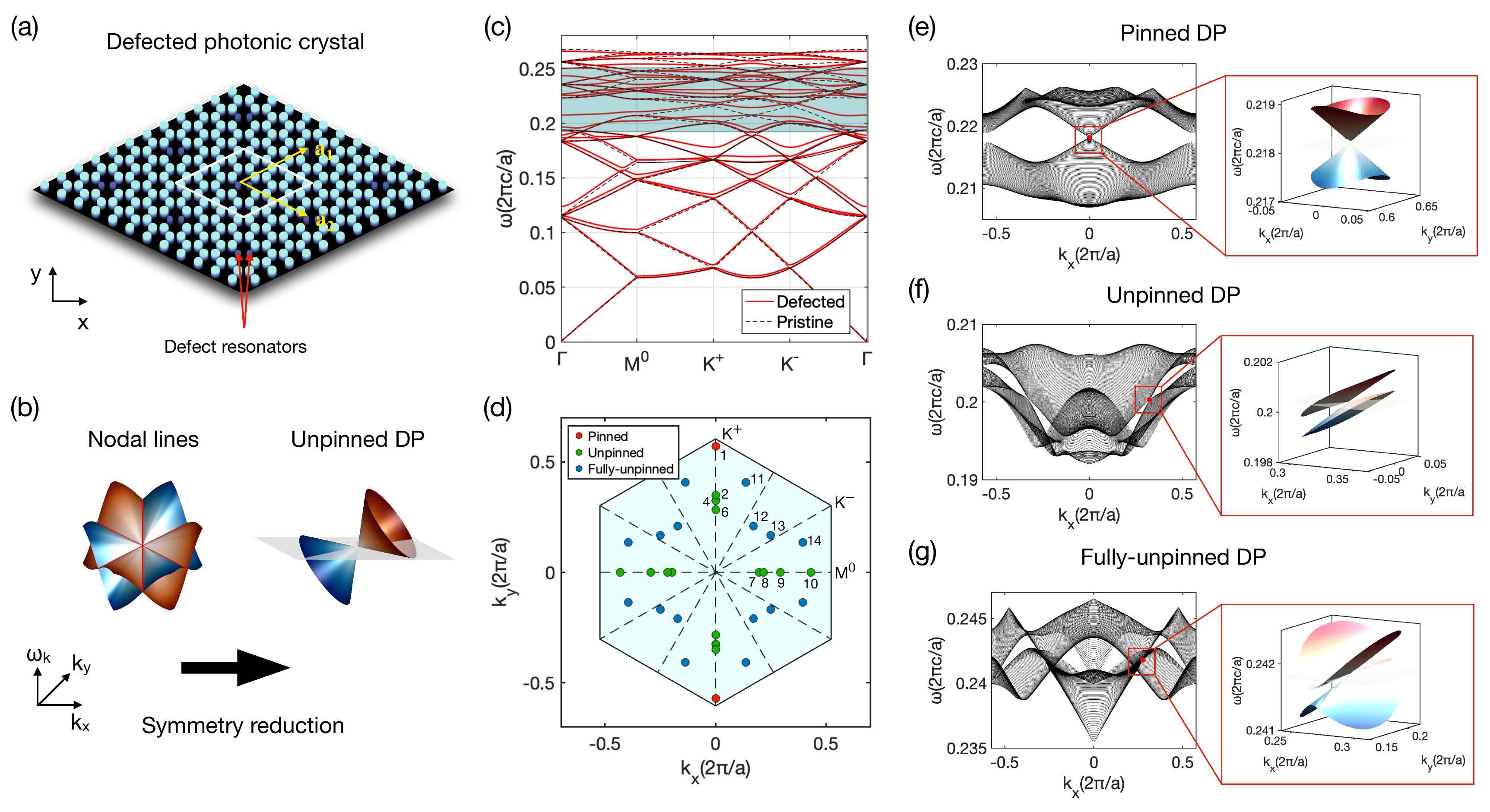}
    \caption{\textbf{Emergence of unpinned Dirac points (DPs).} (a) The unit cell (white line) of the defected photonic crystal (PhC) consists of thirty-two dielectric disks: thirty normal disks (light blue) and two defect disks at the center (blue). The refractive index (\(n\)) or radius (\(R\)) of the defect disks differs from those of the normal disks (\(n_0\) and \(R_0\)), thus creating a superlattice structure. The yellow lines illustrate two lattice vectors (\(\bm{a}_1\) and \(\bm{a}_2\)) of the superlattice. (b) The energy band structure of the pristine PhC, characterized by \((n, R)=(n_0, R_0)\), features nodal line degeneracies along high-symmetry lines (left). Accidental band crossings induced by the symmetry breaking from the defect disks lead to the lifting of these degeneracies, resulting in the emergence of unpinned DPs (right). (c) Thirty-two lowest photonic energy bands of the defected PhC (solid lines) compared to the pristine PhC (dashed lines). The shaded area indicates the frequency range in which DPs are located. (d) Identification of thirty-six DPs from the band structure of the defected PhC in panel (c). The markers indicate the distribution of DPs within the first Brillouin zone (outlined by solid line). The numbers beside each point indicate the labels of the DPs; the omitted numbers ``3" and ``5" overlap with ``2" and ``4," respectively. The dashed lines indicate high-symmetry lines, while the labels \(K^+\), \(K^-\), and \(M^0\) denote the positions of high-symmetry points. (e–g) The sixteenth, fourteenth, and twenty-third bands of the defected PhC, containing (e) a pinned DP (labeled by ``1" in (d)), (f) an unpinned DP (labeled by ``9"), and (g) a fully unpinned DP (labeled by ``13"). The band indices are counted from the lowest to the highest frequency levels. The left images show detailed energy dispersions for each band, with relevant DPs marked by red dots; different curves represent varying \(k_y\) values. The right images display conical energy dispersions around the DPs. The parameter values of \((n, R)=(1,0)\) and \((n_0, R_0)=(4,a/6)\) are applied in panels (c–g). The parameters \(a\) and \(c\) denote the lattice constant of the honeycomb lattice and the speed of light in a vacuum.}
    \label{fig:dirac_point}
\end{figure*}

The photonic superlattice under investigation, as illustrated in Fig.~\ref{fig:dirac_point}(a), is defined by two lattice vectors: \(\bm{a}_1=4a\left(\frac{\sqrt{3}}{2},\frac{1}{2}\right)\) and \(\bm{a}_2=4a\left(\frac{\sqrt{3}}{2},-\frac{1}{2}\right)\), where \(a\) denotes the lattice constant of the underlying honeycomb lattice. The unit cell comprises thirty-two dielectric disks, with two at the center classified as ``defect disks" and the remaining thirty designated as ``normal disks." The defect disks have a radius \(R\) and a refractive index \(n\) that differ from the radius \(R_0\) and refractive index \(n_0\) of the normal disks. The refractive index of the entire system is represented by the piecewise constant function \(n(\bm{r})\), where \(n(\bm{r}) = n_0\) within the normal disks, \(n(\bm{r}) = n\) within the defect disks, and \(n(\bm{r}) = n_{\textrm{air}}=1\) in the surrounding air. In this study, we utilize \(n_0=4\) and \(R_0=\frac{a}{6}\). When \((n, R) = (n_0, R_0)\), the system corresponds to a regular honeycomb lattice structure, referred to as a ``pristine PhC." In contrast, deviations from these parameters indicate the formation of a defected PhC, which exhibits an actual expanded periodicity due to the presence of the defect disks. This deviation breaks three-fold rotational symmetry about the out-of-plane direction, denoted as \(\mathbb{C}_{3z}\), while preserving \(\pi\)-rotational symmetries about the \(x\)- and \(y\)-axes, denoted as \(\mathbb{C}_{2x}\) and \(\mathbb{C}_{2y}\) symmetries, respectively.

We investigate optical modes in a two-dimensional photonic crystal composed of a periodic array of circular dielectric disks. The modes under consideration belong to the transverse magnetic (TM) polarization, where the electric field is oriented in the out-of-plane direction, i.e., $\psi(\bm{r})=(0,0,E_z)$. These modes are determined by solving the Helmholtz equation:
\begin{align}
    -\nabla^2 \psi(\bm{r}) =  n(\bm{r})^2 \frac{\omega^2}{c^2}\psi(\bm{r}), \label{eq:helmholtz}
\end{align}
where $c$ is the speed of light in vacuum, and $\omega = ck$ denotes the free-space temporal frequency with vacuum wavenumber $k$. We numerically solve this equation using the boundary element method with advanced eigenvalue solvers \cite{PhysRevB.54.1880, Jan_2003, Veble_2007, BEM_isakari_bss_fmm, GAO2020102888, IKEGAMI20101927}, imposing dielectric boundary conditions at the disk interfaces and periodic boundary conditions at the supercell boundaries. A detailed description of the method is provided in the Supporting Information (SI) \cite{SI}. Note that the general solution of the Helmholtz equation gives rise to multiple sets of photonic energy bands, each associated with a distinct azimuthal orbital number $l$. In this study, we focus on the fundamental energy bands corresponding to $l = 0$.

The energy band structure of the pristine PhC exhibits degeneracies along high-symmetry lines, such as \(\Gamma\)–\(K^-\) and \(\Gamma\)–\(M^0\), as well as along additional lines connected by rotational symmetry [dashed lines in Fig.~\ref{fig:dirac_point}(c)]. These nodal line structures are typical results of Brillouin zone folding, with their presence and specific positions dictated by the \(\mathbb{C}_{3z}\), \(\mathbb{C}_{2x}\), and \(\mathbb{C}_{2y}\) symmetries of this regular honeycomb lattice. Conversely, the photonic energy bands of the defected PhC lack these nodal line structures, due to the \(\mathbb{C}_{3z}\) symmetry breaking caused by defects, which lifts the degeneracies along the symmetry-protected nodal lines [solid lines in Fig.~\ref{fig:dirac_point}(c)]. Instead, these bands feature thirty-six DPs [Fig.~\ref{fig:dirac_point}(d)], each characterized by a distinct conical dispersion that is non-degenerate, except for an isolated nodal degeneracy point [Fig.~\ref{fig:dirac_point}(e–g)]. Notably, all DPs are positioned away from the typical locations of high-symmetry points, \(K^+\) and \(K^-\) in contrast to the pristine PhC, where two DPs are found at these locations. This observation suggests that accidental band crossings induced by defects are primary factors in the formation of the DPs \cite{Lu2016, Lu2022, doi:10.1142/S2010324716400038}. Furthermore, the DPs are situated at distinct frequency levels (Tab.~\ref{tab:DP_characterization}), particularly in the high-frequency range of \(\omega \gtrsim 0.2 * (2\pi c/a) \) [highlighted by green in Fig.~\ref{fig:dirac_point}(c)]. This phenomenon is attributed to more significant band deformation in the high-frequency bands, while the effect is comparatively weaker in the low-frequency bands, a characteristic associated with the short-range nature of the defects.

We classify the observed DPs regarding their symmetry properties. Two of these DPs correspond to typical DPs found in a honeycomb lattice and are referred to as ``pinned DPs," with their persistence attributed to retained sublattice symmetry \cite{RevModPhys.81.109}. The remaining thirty-four points correspond to newly identified DPs that arise from the arrangement of defected disks. Eighteen of these points align along high-symmetry lines, such as the \(k_x\) and \(k_y\) axes, with each pair connected by either \(\mathbb{C}_{2x}\) or \(\mathbb{C}_{2y}\) symmetry. The other sixteen points are situated at generic \(k\) points that do not align with any high-symmetry points or lines. These points form groups of four related by both \(\mathbb{C}_{2x}\) and \(\mathbb{C}_{2y}\) symmetries. The former and latter groups are designated as ``unpinned DPs" and ``fully unpinned DPs," respectively, and will be discussed further. In total, there are fourteen distinct DPs: one from the pinned DPs, nine from the unpinned DPs, and four from the fully unpinned DPs. Each of these DPs is independent and exhibits distinct energy dispersions. The dispersions of three DPs from each group are shown in Fig.~\ref{fig:dirac_point}(e–g). The configurations of the remaining eleven DPs, along with further details on all fourteen DPs, can be found in SI \cite{SI}.

To characterize the fourteen distinct DPs, we employ the \(k\cdot p\) theory \cite{PhysRevB.86.035141, Wang2017, PhysRevLett.121.024301}, which describes the energy dispersion around a DP using an effective two-band model:
\begin{equation}
    H_\textrm{Dirac}(\bm{q}) = (\omega_0 + \bm{u} \cdot \bm{q}) \hat{I} + v_1 q_1 \sigma_1 + v_2 q_2 \sigma_2. \label{eq:H_dirac}
\end{equation}
In this equation, \(\hat{I}\) is the two-by-two identity matrix, and \(\sigma_1\) and \(\sigma_2\) are the Pauli matrices. \(\omega_0\) denotes the frequency at a nodal point. The momentum vector \(\bm{q} = \bm{k} - \bm{k}_*\) represents the deviation of the actual momentum \(\bm{k}\) from the position of a DP \(\bm{k}_*\). It is expressed as \(\bm{q} = q_1\hat{\bm{e}}_1 + q_2\hat{\bm{e}}_2\), where \(\hat{\bm{e}}_1\) and \(\hat{\bm{e}}_2\) denote two orthogonal unit vectors pointing the directions of the steepest and least energy dispersion within a local momentum region of interest, respectively. The two eigenenergies of \(H_\textrm{Dirac}(\bm{q})\) are represented as
\begin{equation}
    \omega_{\pm}(\bm{q}) = \omega_0 + \bm{u} \cdot \bm{q} \pm \sqrt{(v_1 q_1)^2 + (v_2 q_2)^2}, \label{eq:omega_q}
\end{equation}
where the signs \(+\) and \(-\) correspond to the upper and lower bands, respectively. The two functions \(\omega_{\pm}(\bm{q})\) represent a general conical dispersion with potential anisotropic features, associated with a DP. The parameters \(v_1\) and \(v_2\) represent the ``relative velocities" of the two bands in the \(\hat{\bm{e}}_1\) and \(\hat{\bm{e}}_2\) directions, respectively. Meanwhile, the parameters \(u_1\) and \(u_2\) represent the ``average velocities" of the two bands in the same directions, where \(\bm{u} = u_1\hat{\bm{e}}_1 + u_2\hat{\bm{e}}_2\) and the reverse vector \(-\bm{u}\) indicates the direction of the tilt of the Dirac cone. With respect to the four parameters \((v_1, v_2, u_1, u_2)\), two characteristic quantities that describe Dirac energy dispersions are defined as follows:
\begin{equation}
    \mu = \sqrt{\left(\frac{u_1}{v_1}\right)^2 + \left(\frac{u_2}{v_2}\right)^2}, ~ \eta = \frac{v_1}{v_2}. \label{eq:cone_params}
\end{equation}
The tilt parameter (\(\mu\)) defines the degree of tilt of each Dirac cone \cite{PhysRevB.86.035141, Wang2017, PhysRevLett.121.024301}, and the anisotropy parameter (\(\eta\)) quantifies the anisotropy of the Dirac cone along the \(\hat{\bm{e}}_1\) and \(\hat{\bm{e}}_2\) directions. The four parameters \((v_1, v_2, u_1, u_2)\) for each DP are estimated by fitting the two relevant energy bands to the dispersions described in Eq.~\eqref{eq:omega_q}. For a detailed description of the fitting procedure, refer to SI \cite{SI}. 

\begin{table}[]
    \centering
    \begin{tabular}{p{1.05cm}|p{1.05cm}p{1.05cm}p{1.05cm}p{1.05cm}p{1.05cm}p{1.05cm}p{1.05cm}p{1.05cm}p{1.05cm}p{1.05cm}p{1.05cm}p{1.05cm}p{1.05cm}p{1.05cm}p{1.05cm}} \toprule
        DP no. & 1 & 2 & 3 & 4 & 5 & 6 & 7 & 8 & 9 & 10 & 11 & 12 & 13 & 14 \\ \midrule
        \(\omega_0\) & 0.218 & 0.247 & 0.235 & 0.199 & 0.226 & 0.229 & 0.230 & 0.241 & 0.200 & 0.246 & 0.232 & 0.250 & 0.242 & 0.239 \\
        \(\mu\) & 0.079 & 0.343 & 3.739 & 2.270 & 1.228 & 1.006 & 0.368 & 0.137 & 2.580 & 0.079 & 0.752 & 1.999 & 1.811 & 1.306 \\
        \(\eta\) & 1.334 & 1.916 & 6.658 & 1.909 & 1.827 & 2.030 & 4.198 & 6.237 & 1.199 & 4.129 & 3.454 & 2.563 & 4.034 & 5.014 \\
        Type & P/I & U/I & U/II & U/II & U/II & U/II & U/I & U/I & U/II & U/I & FU/I & FU/II & FU/II & FU/II \\ \bottomrule 
    \end{tabular}
    \caption{\textbf{Characterization of the fourteen distinct DPs.} The numbers in ``DP no." refer to the label number of each DP, as presented in Fig.~\ref{fig:dirac_point}(d). The frequency level at each DP (\(\omega_0\)) is expressed in units of \(2\pi c/a\). The parameters \(\mu\) and \(\eta\) represent the tilt and anisotropy of the conical dispersion, as defined in Eq.~\eqref{eq:cone_params}. The letters ``P," ``U," and ``FU" signify ``pinned DP," ``Unpinned DP," and ``fully unpinned DP," respectively, while ``I" and ``II" indicate type-I (\(\mu < 1\)) and type-II (\(\mu \geq 1\)), respectively. }
    \label{tab:DP_characterization}
\end{table}

Table~\ref{tab:DP_characterization} presents the values of \(\mu\) and \(\eta\) for the fourteen distinct DPs. The pinned DP has \((\mu, \eta) \approx (0.1, 1.3)\), which is close to \((\mu,\eta)=(0,1)\) of a typical pinned DP in the pristine PhC. Consequently, the pinned DP exhibits a nearly non-tilted, isotropic conical dispersion despite the symmetry-breaking effects of defects [Fig.~\ref{fig:dirac_point}(e)]. On the other hand, the unpinned and fully unpinned DPs have significant values of \(\mu\) or \(\eta\), exhibiting tilted, anisotropic conical dispersions [Fig.~\ref{fig:dirac_point}(f–g)]. Such characteristics are unprecedented in typical DPs \cite{PhysRevLett.100.013904, PhysRevA.78.033834, PhysRevA.75.063813, PhysRevLett.100.113903, DIEM20102990, doi:10.1073/pnas.1207335109, PhysRevB.86.035141, Li:13, https://doi.org/10.1002/lpor.201300186, PhysRevA.96.013813, PhysRevLett.121.263901, Navarro-Barón2021}, and represent unique features of the unpinned and fully unpinned DPs. Based on the \(\mu\) values, the fourteen distinct DPs are grouped into two categories: type-I for \(\mu < 1\) and type-II for \(\mu \geq 1\) \cite{PhysRevB.86.035141, Wang2017, PhysRevLett.121.024301}. The type-I DP features a single intersection point on the equal-frequency surface at the frequency level of the DP [Fig.~\ref{fig:dirac_point}(e)]. In contrast, the type-II DPs are characterized by two intersecting lines on the equal-frequency surface [Fig.~\ref{fig:dirac_point}(f–g)]. The numbers of type-I and type-II DPs in the defected PhC are found to be \(N_\textrm{I} = 6\) and \(N_\textrm{II} = 8\), respectively. Within each type, the fourteen distinct DPs can be further grouped into two categories: one exhibiting isotropic dispersions with \(\eta \approx 1\) and the other exhibiting anisotropic dispersions with \(\eta \gg 1\). Consequently, the defected PhC displays all possible categories of conical dispersions: (i) non-tilted, isotropic dispersions \((\mu\approx0, \eta\approx1)\), (ii) tilted, isotropic dispersions \((\mu\approx0, \eta\approx1)\), (iii) non-tilted, anisotropic dispersions \((\mu\approx0, \eta\gg1)\), and (iv) tilted, anisotropic dispersions \((\mu>1, \eta\gg1)\) (Tab.~\ref{tab:DP_characterization}). This diversity in conical dispersions suggests that the defected PhC can serve as a versatile platform for the exploration of relativistic photonic wave phenomena with greater flexibility, compared to conventional PhCs.

The emergence of thirty-six DPs in the defected PhC is noteworthy, especially considering that only two DPs exist in the pristine PhC. This abundance of DPs may arise from the numerous symmetry-protected nodal line degeneracies present in the pristine PhC, which can transform into isolated nodal point degeneracies, specifically DPs, through accidental band crossings induced by symmetry breaking. However, this reasoning is not straightforward, as avoided crossings are generally prevalent and hinder the formation of isolated nodal points in the absence of symmetry constraints \cite{qc0,qc1,mcavity0}. Thus, there must be underlying physical mechanisms that prevent avoided crossings. We posit that the preserved \(\mathbb{C}_{2x}\) and \(\mathbb{C}_{2y}\) symmetries are crucial for the stability of these unpinned DPs against avoided crossings. This situation is reminiscent of sublattice symmetry, which prevents the opening of a bandgap at DPs in graphene \cite{RevModPhys.81.109}. To validate this hypothesis, we investigated a defected PhC, where the \(\mathbb{C}_{2y}\) symmetry is additionally broken due to the inequivalence of the two defect disks; for a detailed analysis, refer to SI \cite{SI}. Our findings indicate that this symmetry breaking leads to gap openings at certain DPs, thereby supporting our assertion. Nevertheless, identifying the precise physical role of this symmetry in preventing avoided crossings remains an open problem.
\begin{figure*}[t!]
    \centering
    \includegraphics[width=\textwidth]{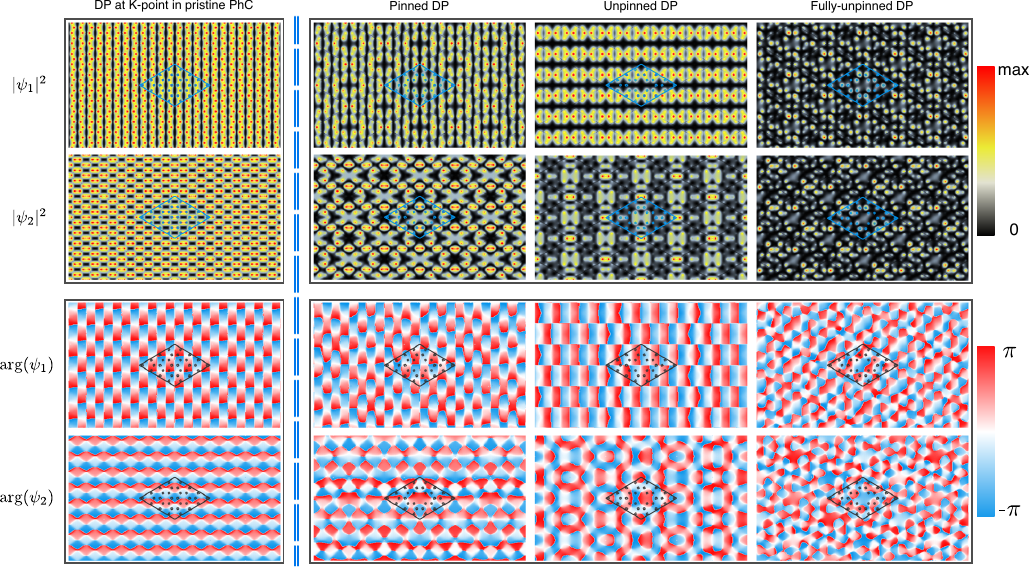}
    \caption{\textbf{Photonic wavefunctions of DPs in the pristine and defected PhCs.} The wavefunctions, denoted as $\psi_{1,2} = (0,0,E_z)$, represent the electric field oriented in the out-of-plane direction [i.e., transverse magnetic (TM) polarization], where $\psi_1$ and $\psi_2$ correspond to two degenerate photonic modes at each Dirac point (DP). The first column shows \(\psi_1\) and \(\psi_2\) at the pinned DP in the pristine PhC, while the second, third, and fourth columns illustrate \(\psi_1\) and \(\psi_2\) at the pinned, unpinned, and fully unpinned DPs in the defected PhC, respectively, as presented in Fig.~\ref{fig:dirac_point}(e–g). In each column, the first and second rows present the wavefunction intensities (\(|\psi_1|^2\) and \(|\psi_2|^2\)), while the third and fourth rows indicate their phases [\(\textrm{arg}(\psi_1)\) and \(\textrm{arg}(\psi_2)\)]. In each panel, the blue or black circles represent the disk sites, while the lines in the same colors denote the unit cell.}
    \label{fig:wavefunctions}
\end{figure*}

The photonic wavefunctions for two degenerate modes at the pinned DP in the pristine PhC exhibit odd or even parity for \(\mathbb{C}_{2y}\), attributed to the high-symmetry nature of the \(K^+\) point [see \(\textrm{arg}(\psi_1)\) and \(\textrm{arg}(\psi_2)\) in the first column of Fig.~\ref{fig:wavefunctions}]. Similar characteristics are observed in both pinned and unpinned DPs in the defected PhC, where the wavefunctions also display odd or even parity with respect to either \(\mathbb{C}{2x}\) or \(\mathbb{C}_{2y}\), depending on their momentum locations [\(\textrm{arg}(\psi_1)\) and \(\textrm{arg}(\psi_2)\) in the second and third columns]. In contrast, the fully unpinned DP in the defected PhC lacks any definite parity, attributed to its asymmetric location in momentum space [\(\textrm{arg}(\psi_1)\) and \(\textrm{arg}(\psi_2)\) in the fourth column]. Meanwhile, the pinned DP in the defected PhC exhibits a slight transfer of the wavefunction intensity from the defect sites to adjacent normal sites. In contrast, the wavefunctions for the unpinned and fully unpinned DPs tend to display highly concentrated intensity at specific sites (\(|\psi_1|^2\) and \(|\psi_2|^2\) in the second to and fourth columns of Fig.~\ref{fig:wavefunctions}), compared to those for the DP in the pristine or defected PhCs (\(|\psi_1|^2\) and \(|\psi_2|^2\) in the first and second columns). Our calculation of the inverse participation ratio (IPR) \cite{PhysRevB.83.184206}, defined as \( \frac{\int dx \int dy |\psi_{1,2}(x,y)|^4}{\int dx \int dy |\psi_{1,2}(x,y)|^2},\) reveals that the IPR values for unpinned and fully unpinned DPs are up to 2.6 times greater than those of the DP in the pristine PhC. Consequently, we conclude that the wavefunctions of the unpinned and fully unpinned DPs in the defected PhC are characterized by either a lack of symmetry properties or enhanced wavefunction intensities, significantly differing from those of the conventional pinned DP. For a detailed analysis of the IPR, refer to SI \cite{SI}.

\subsection{Defect engineering of unpinned Dirac points}

\begin{figure*}[t!]
    \centering
    \includegraphics[width=.9\textwidth]{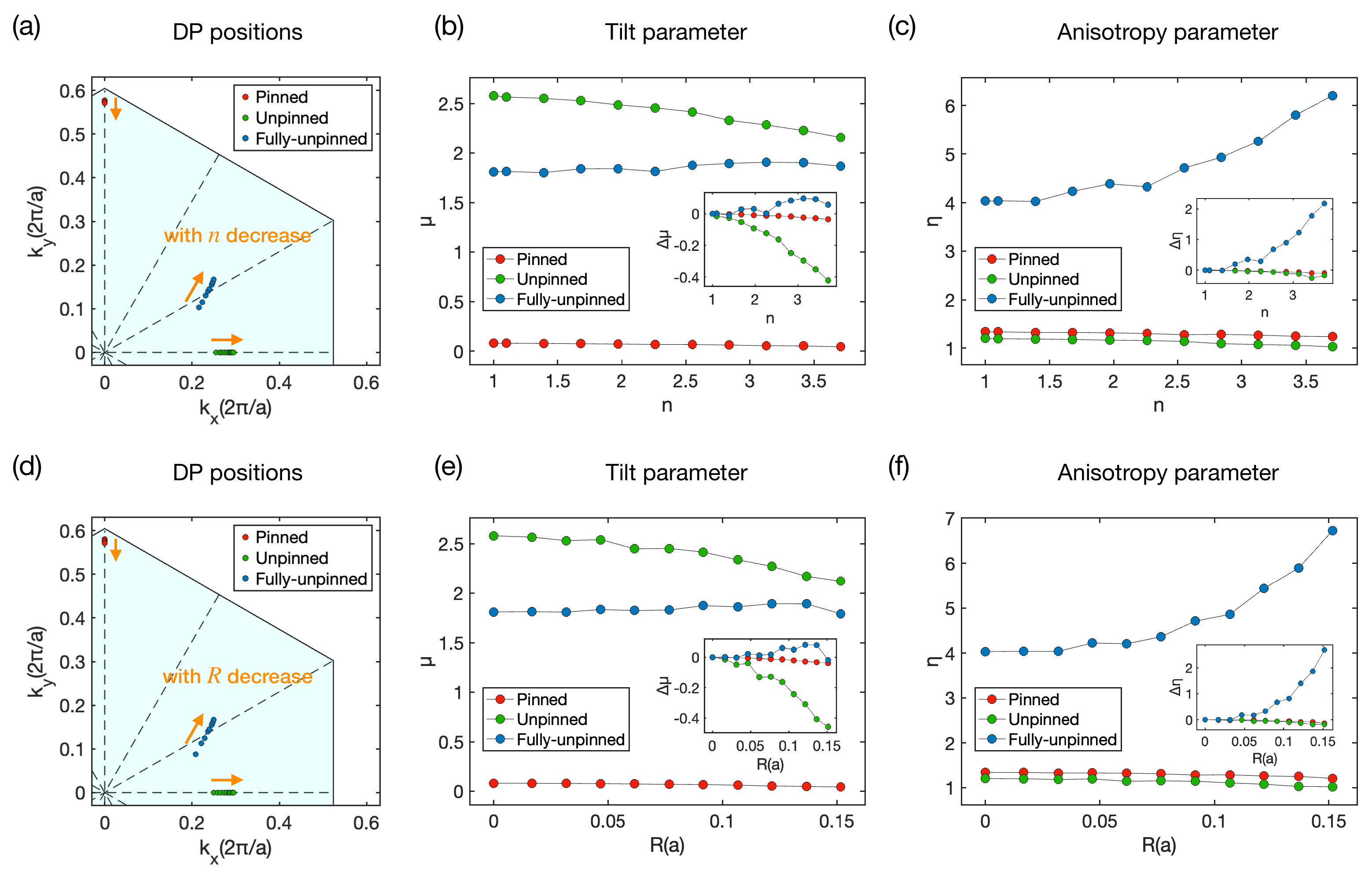}
    \caption{\textbf{Influence of adjustments to defect disks on unpinned DPs.} (a) Shifts of three DPs, labeled by ``Pinned," ``Unpinned," and ``fully unpinned," with respect to the variation in \(n\). The three DPs correspond to those depicted in Fig.~\ref{fig:dirac_point}(e–g). The markers indicate the location of the three DPs, where the yellow arrows guide the direction of the shifts of the DP positions with the decrease of \(n\). The solid line outlines the first Brillouin zone in the first quadrant, while the dashed lines indicate high-symmetry lines. (b–c) Variation in (b) \(\mu\) and (c) \(\eta\) of the DPs as a function of \(n\). The insets show the deviations defined as \(\Delta\mu(n) = \mu(n) - \mu(1)\) and \(\Delta\eta(n) = \eta(n) - \eta(1)\). (d) Shifts of the three DPs with respect to the variation in \(R\). The markers indicate the location of the three DPs, where the yellow arrows guide the direction of the shifts of the DP positions with the decrease of \(R\). (e–f) Variation in (e) \(\mu\) and (f) \(\eta\) of the DPs as a function of \(R\). The insets show the deviations defined as \(\Delta\mu(R) = \mu(R) - \mu(0)\) and \(\Delta\eta(R) = \eta(R) - \eta(0)\). In panels (a–c), the parameter \(R\) is set to \(0.092a\), whereas in panels (d–f), \(n\) is set to \(2.55\). In all panels, the parameter values \((n_0, R_0)=(4,a/6)\) are utilized. }
    \label{fig:dirac_dispersion}
\end{figure*} 

We investigate how the values of \((n, R)\) of defect disks influence the properties of DPs, with a specific focus on the three DPs, as shown in Fig.~\ref{fig:dirac_point}(d). The deviations of \((n, R)\) from the reference values \((n_0, R_0)=(4,a/6)\) of the normal disk are defined as follows:
\begin{equation}
    \delta n = 4 - n, ~\delta R = \frac{a}{6} - R.
\end{equation}
At \((\delta n, \delta R) = (0, 0)\), corresponding to the pristine PhC, the pinned DP is positioned at \(K^+\) due to the \(\mathbb{C}_{3z}\) symmetry. As \(\delta n\) increases, the pinned DP shifts away from \(K^+\) along a high-symmetry line \(\Gamma\)–\(K^+\), with the magnitude of this shift increasing in proportion to the deviation \(\delta n\) [red points in Fig.~\ref{fig:dirac_dispersion}(a)]. However, even for the largest deviation of \(\delta n_\textrm{max} = 3 \), the magnitude of the shift remains small, keeping the pinned DP close to \(K^+\). This behavior aligns with prior observations in graphene nanomeshes that lack \(\mathbb{C}_{3z}\) symmetry \cite{Jia2016}. This indicates that while the original \(\mathbb{C}_{3}\) symmetry is ``weakly" broken by the defect disks, it still imposes significant constraints on this DP. The symmetry origin and observed pinned nature reinforce our prior designation of this DP as a ``pinned DP" \cite{doi:10.1142/S2010324716400038}. Additionally, the pinned DP exhibit minimal variation in both \(\mu\) and \(\eta\) with changes in \(\delta n\), maintaining values of \(\mu \approx 0\) and \(\eta \approx 1.3\) at the optimal deviation \(\delta n_\textrm{max}\) [red points in Fig.~\ref{fig:dirac_dispersion}(b–c)]. Similar behavior is also observed with \(R\); the pinned DP shifts away from \(K^+\) in proportion to the deviation \(\delta R\) [red points in Fig.~\ref{fig:dirac_dispersion}(d)] and exhibits uniform values of \(\mu\) and \(\eta\) with respect to \(\delta R\) [red points in Fig.~\ref{fig:dirac_dispersion}(d–f)].

As \(\delta n\) or \(\delta R\) increases, the unpinned DP shifts along the high-symmetry line \(\Gamma\)–\(M^0\) [green points in Fig.~\ref{fig:dirac_dispersion}(a, d)]. In contrast, the fully unpinned DP moves to generic momentum points, irrespective of high-symmetry lines or points [blue points in Fig.~\ref{fig:dirac_dispersion}(a, d)]. In both DPs, the magnitudes of the shifts are proportional to \(\delta n\) and \(\delta R\) and are significant compared to the pinned DP. These mobility characteristics are consistent with the known behaviors of unpinned and fully unpinned DPs \cite{Lu2016, Lu2022, doi:10.1142/S2010324716400038}, reinforcing our previous classifications as ``unpinned" and ``fully unpinned" DPs, respectively. Furthermore, the unpinned DPs exhibit a significant increase in \(\mu\) in response to decrease in \(n\) or \(R\) [green points in Fig.~\ref{fig:dirac_dispersion}(b–c) and Fig.~\ref{fig:dirac_dispersion}(e–f)], while the fully unpinned DP shows a substantial decrease in \(\eta\) with respect to decrease in \(n\) or \( R\) [blue points in Fig.~\ref{fig:dirac_dispersion}(b–c) and Fig.~\ref{fig:dirac_dispersion}(e–f)]. We observe similar mobility characteristics and substantial changes in \(\mu\) or \(\eta\) among the remaining eleven DPs; however, detailed patterns, such as increasing or decreasing trends in \(\mu\) or \(\eta\) related to variations in \(n\) and \(R\), differ for each DP. For further details, refer to SI \cite{SI}. These observations lead us to conclude that unpinned and fully unpinned DPs demonstrate greater tunability in response to changes in these values—a unique feature not observed in typical DPs, such as the pinned DP in our investigation.

We emphasize that most DPs, including the three investigated here, persist regardless of the specific values of \(n\) or \(R\). In contrast, two DPs labeled 8 and 14 may no longer be identifiable when \(n\) is either large or small, or when \(R\) is either large or small, since they exhibit significant deviations from typical conical dispersions; further details can be found in SI \cite{SI}. Given that accidental band crossings are typically sensitive to the specifics of the system, the observed robustness of the DPs is noteworthy. We note that the \(\mathbb{C}_{2x}\) and \(\mathbb{C}_{2y}\) symmetries are maintained across all variations of defected PhCs regardless of \(n\) or \(R\), suggesting that the preservation of these symmetries plays a crucial role in the emergence of unpinned DPs, aligned our previous analysis. Consequently, this suggests that unpinned and fully unpinned DPs can be achieved without fine-tuning specific parameters such as the refractive index and size of the defect disks (\(n\) and \(R\)).

\subsection{Dirac photon phase diagram}

\begin{figure*}[t!]
    \centering
    \includegraphics[width=\textwidth]{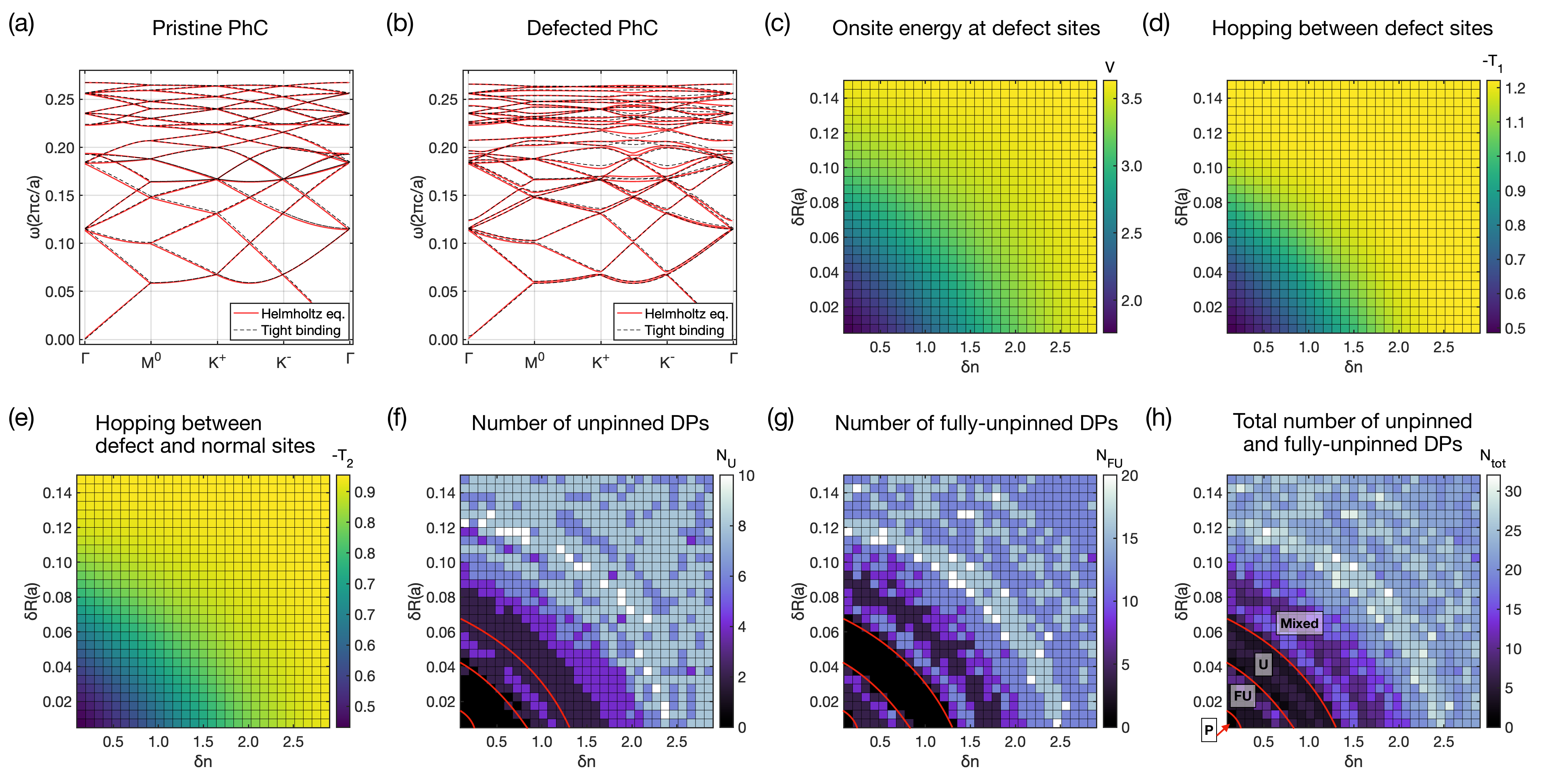}
    \caption{\textbf{Phase diagram derived from the effective tight-binding model.} (a–b) The energy bands derived from the effective photonic tight-binding model (solid line) are compared to those obtained from the Helmholtz equation (dashed line). Panel (a) corresponds to the pristine PhC, while panel (b) corresponds to the defected PhC characterized by \((n, R)=(1,0)\). (c–e) Three principal matrix elements in the tight-binding model exhibiting significant alternation with \(\delta n = n_0 - n\) and \(\delta R = R_0 - R\). Each panel depicts: (c) the onsite energy at the defect disks (\(V\)), (d) the hopping amplitude between the two defect disks (\(T_1\)), and (e) the hopping amplitude between a defect disk and its neighboring normal disk (\(T_2\)) as a function of \((\delta n, \delta R)\). (f–h) Phase diagrams for the defected PhC, illustrating the number of unpinned (\(N_{U}\)) and fully-pinned DPs (\(N_{FU}\)), derived from the effective tight-binding model. (f) shows \(N_{U}\), (g) shows \(N_{FU}\), and (h) presents their sum (\(N_{tot}=N_{U}+N_{FU}\)). The number of pinned DPs remains fixed at two across all phases. Four distinct phases are identified based on the values of \(N_U\) and \(N_{FU}\) as follows: (i) pinned DP phase (\(N_{U}=N_{FU}=0\)), (ii) fully unpinned DP phase (\(N_{U}=0, N_{FU}>0\)), (iii) unpinned DP phase (\(N_{U}>0, N_{FU}=0\)), and (iv) mixed DP phase (\(N_{U}, N_{FU}>0\)). These phases are labeled as ``P", ``FU", ``U", and ``Mixed" in (h), respectively. Red lines indicate phase boundaries.}
    \label{fig:phase_diagram}
\end{figure*} 

We employ the effective tight-binding model framework to simulate the energy bands of the defected PhC \cite{Yi2022, Yi2023}. This method enables us to avoid resource-intensive simulations involving various combinations of \(n\) and \(R\) and circumvent significant numerical challenges, such as the emergence of spurious solutions \cite{spurious0,spurious1,spurious2,spurious3}. It is particularly suitable for the tunneling regime of interest, which is characterized by a high refractive index \((n_0 \gg n_{\text{air}})\) or a small disk size \((R_0 \ll a)\) \cite{Yi2022, Yi2023}. The tight-binding model is formulated as follows:
\begin{equation}
    H = \sum_{i} v_i | \bm{r}_i \rangle \langle \bm{r}_i | + \sum_{i,j} (t_{ij} | \bm{r}_i \rangle \langle \bm{r}_j | + \text{h.c.}). \label{eq:tb_model}
\end{equation}
In this equation, the state \(|\bm{r}_i \rangle\) denotes a ``Wannier wavefunction" at the \(i\)-th site, with \(v_i\) representing the onsite potential energy at that site and \(t_{ij}\) indicating the hopping amplitudes between the \(i\)-th and \(j\)-th sites. The values of \(v_i\) and \(t_{ij}\) are derived from Bloch wavefunctions reconstructed from the solution of the Helmholtz equation along with their corresponding energy levels. We derive tight-binding models for the pristine and defected PhCs incorporating various combinations of \((\delta n, \delta R)\). The effective photonic energy bands are derived from the square roots of the eigenvalues of these models, which accurately reproduce the energy bands calculated from the Helmholtz equation, as illustrated in Fig.~\ref{fig:phase_diagram}(a–b). Furthermore, the DPs derived from the tight-binding models exhibit characteristics consistent with those in the defected PhC, confirming the validity of our approach. For a detailed analysis of these DPs and a description of the numerical algorithm, refer to SI \cite{SI}.

The tight-binding model for the defected PhC exhibits site-to-site variations in both \(v_i\) and \(t_{ij}\) due to translational symmetry breaking induced by the defect disks, in contrast to the pristine PhC. Three parameters in \(v_i\) and \(t_{ij}\) are directly linked to the defect disks: the site energy \(v_i\) at the defect disks (denoted as \(V\)), the coupling \(t_{ij}\) between two defect disks (\(T_1\)), and the coupling \(t_{ij}\) between a defect disk and a neighboring normal disk (\(T_2\)). These parameters significantly increase in a symmetric manner with both \(\delta n\) and \(\delta R\) [Fig.~\ref{fig:phase_diagram}(c–e)]. Specifically, \(V\) increases to 3.6, \(T_1\) increases to 1.2, and \(T_2\) increases to 0.9, representing multipliers of two, three, and two times greater than the pristine values of \(V \approx 1.6\), \(T_1 \approx 0.4\), and \(T_2 \approx 0.4\). Physically, these parameters indicate that the photonic wavefunction is ``less" occupied at the defect disks due to their reduced refractive index or size compared to the normal disks. In contrast, the other elements, including the site energy at the normal disks and the hopping amplitudes between two nearest normal disks, show only slight variations, since they are not associated with the defect disks; detailed presentations of these parameters can be found in SI \cite{SI}. Consequently, the defected PhC can be effectively represented by a tight-binding model with augmented onsite potentials and hopping terms (\(V\), \(T_1\), and \(T_2\)).

Using the developed tight-binding models, we establish global phase diagrams for the defected PhC, illustrating the distribution of unpinned and fully unpinned DPs as functions of \((\delta n, \delta R)\), as shown in Fig.~\ref{fig:phase_diagram}(f–g). These diagrams reveal four distinct phases: (i) pinned, (ii) fully unpinned, (iii) unpinned, and (iv) mixed DP phases. In the pinned DP phase, where \((\delta n, \delta R) \approx (0, 0)\), only two pinned DPs are present, and neither unpinned nor fully unpinned DPs appear in the energy bands. As \((\delta n, \delta R)\) exceed certain threshold values, the system transitions into the fully unpinned DP phase, where fully unpinned DPs emerge, while unpinned DPs remain absent. Conversely, with further increases in \(\delta n\) or \(\delta R\), the system shifts into the unpinned DP phase, where fully unpinned DPs disappear and unpinned DPs arise. Finally, in the mixed DP phase, observed at significant values of \((\delta n, \delta R)\), both unpinned and fully unpinned DPs are present. The quantities of both unpinned and fully unpinned DPs tend to increase with \(\delta n\) and \(\delta R\), exhibiting symmetry in their behavior, which implies their comparable roles in the emergence of these DPs. Furthermore, this trend aligns with the increasing patterns in \(-T_1\), \(-T_2\), and \(V\), as illustrated in Fig.~\ref{fig:phase_diagram}(c–e), suggesting their critical contributions to the formation of these DPs. Consequently, we conclude that defected PhCs can accommodate various combinations of pinned, unpinned, and fully unpinned DPs by adjusting the refractive index or size of the dielectric disks, as denoted by \(\delta n\) and \(\delta R\) in our theoretical treatment.

\section{Discussion}

We have conducted the first investigation into the potential of defected PhCs, a novel structure inspired by porous nanomeshes in 2D materials \cite{Bai2010, C4NR04584J}. This study has revealed the emergence of unpinned DPs, a phenomenon that has been insufficiently explored within the realm of PhCs. Unlike the pinned DPs identified earlier, these unpinned DPs exhibit significant tunability in their quantity, momentum-space locations, and velocities. The proposed defected PhC can be experimentally realized with silicon-based materials, requiring a wavelength in the range of \(0.35 \mu m < \lambda < 2 \mu m\) to achieve a refractive index of \(3.5 < n < 7\). In semiconductor alloys such as GaAs, AlGaSb, and InGaAsP, a similar wavelength range gives rise to a refractive index of \(3.3 < n < 5\). Additionally, the mixed structure composed of normal and defect disks can be created using modern state-of-the-art etching technologies such as those given in \cite{etching0,etching1}. The presence of Dirac points can be identified as nodal point degeneracies through transmission measurements \cite{Wang2019, doi:10.1126/science.aaq1221, Noh2017, https://doi.org/10.1002/lpor.201700271, https://doi.org/10.1002/lpor.202100452}. Additionally, observing anomalous refraction and giant birefringence serves as reliable indicators for distinguishing between type-I and type-II Dirac points, offering detailed information into their dispersion characteristics \cite{Wang2017, Wang:22}.

We emphasize that our defect engineering approach extends beyond the PhC in a honeycomb lattice examined in this study; it is applicable to various physical systems with different lattice structures. This versatility encompasses phononic crystals and optical lattices, where the required mixed structure composed of normal and defect elements can be created with precise control. Additionally, this method may be adapted to solid-state materials, where required defect elements can be introduced by the strategic removal of atoms at designated sites through an electron beam or optical lithography \cite{Bai2010, C4NR04584J}. For this pursuit, further large-scale theoretical research will likely be required, as current lithography technologies have a resolution limited to dozens of atomic sites, which is significantly larger than the present consideration. Importantly, investigating the relativistic physics governed by the Dirac equation in these inhomogeneous superlattice systems presents a new mathematical challenge: developing a comprehensive effective tight-binding model and understanding the algebraic structure of relevant matrices, which are large and composed of non-uniform elements. Addressing this challenge could open significant future research avenues, including the discovery of novel phenomena associated with DPs beyond conventional material platforms.

\section*{Data availability}

The data provided in the manuscript is available from K.K. upon request.

\section*{ACKNOWLEDGEMENTS}

We extend our gratitude to Jung-Wan Ryu, Beom Hyun Kim, and Hee Chul Park for sharing their valuable insights. C.-H. Yi and K.-M.K. received support from the Institute for Basic Science in the Republic of Korea through the project IBS-R024-D1. K.-M.K. was also supported by an appointment to the JRG Program at the APCTP through the Science and Technology Promotion Fund and Lottery Fund of the Korean Government. This research was also supported by the Korean Local Governments - Gyeongsangbuk-do Province and Pohang City. \\

\bibliography{ref}

\end{document}